\documentstyle[epsfig]{bpl}
\def \pt {p_{\rm T}}

\begin{document}
\setcounter{page}{1}

~\\
\bpl  27-31 October 2008\\
BPL, {\bf 8} (2), pp. 1 - 4  (2008) \vskip14mm

\title{W and Z boson Productions in CMS at LHC startup}
%%\title{%%Production of jets accompanied by W and Z bosons at LHC startup}

\author{Didar Dobur \\Istituto Nucleare di Fisica Nazionale(INFN)
Pisa, Italy,\\Currently at the University of Florida, Gainsville, USA.\\[5mm]}

\maketitle

\centerline{(15 January 2009)}

\abstract{We report on potential for measurement of W and Z boson production, as well as the production in association with
jets. 
Of particular interest are jet multiplicity and $P_{\rm T}$ distributions. The 10 to $100~{\rm pb}^{-1}$ 
datasets expected in the startup year of operation of LHC are likely to already provide 
information beyond the reach of the Tevatron collider both in jet multiplicity and $P_{\rm T}$ range. 
We are especially interested in understanding the ratios of W+jets to Z+jets distributions by 
comparing them to next-to-leading order Monte Carlo generators, as these processes present a 
formidable background for searches of new physics phenomena.}\ea

\section{W(Z) production in CMS}

The decays of W and Z bosons into leptons provide a clean experimental signature, and 
the large production cross section for W(Z) boson, i.e. 190 ${\rm nb}$(56 ${\rm nb}$) at $\sqrt{s}=14$ TeV
allows to measure these processes with the very early data of LHC. Since the properties of W(Z) bosons are relatively
well known, these measurements will be extremely important for understanding of detector performances. 

\begin{figure}
\begin{center}
  \includegraphics[scale=0.30]{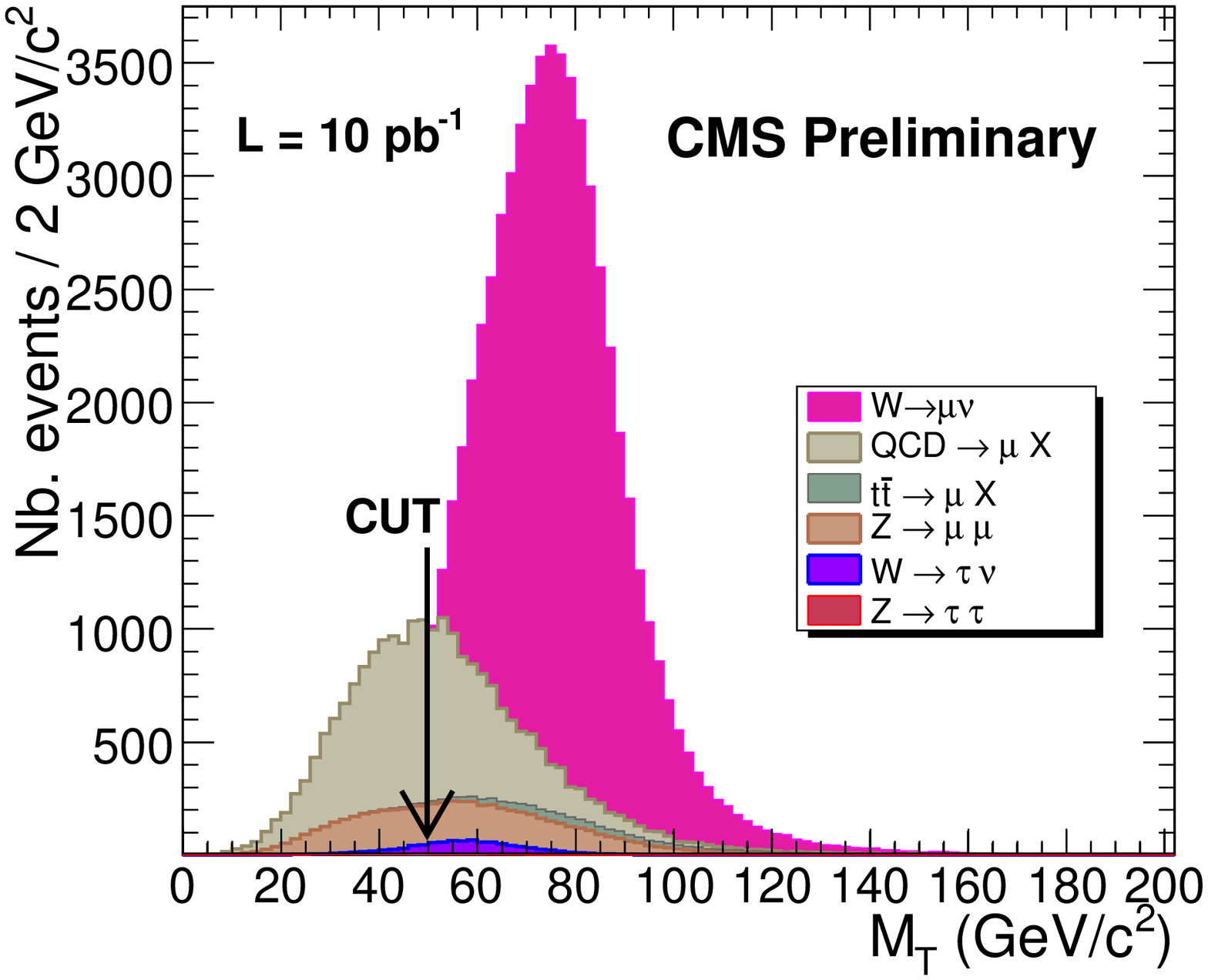}
  \includegraphics[scale=0.30]{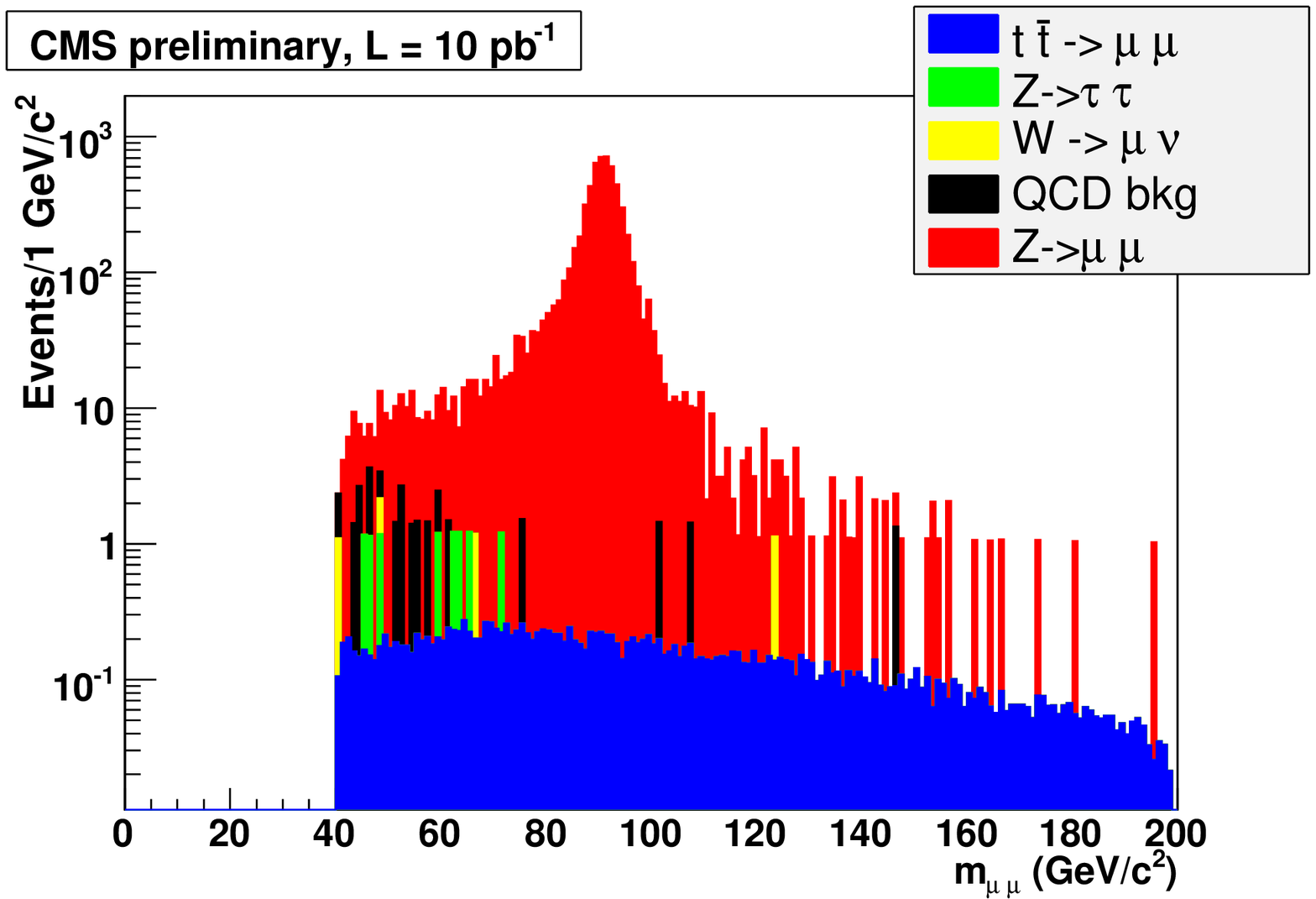} 
  \caption{Left: Transverse mass, $M_{\rm T}$, of the selected W 
  candidates from signal and the background samples, Right: di-muon invariant mass, $M_{\mu \mu}$, for selected Z
  candidates from signal and background samples, for an integrated luminosity of $10~pb^{-1}$.}
  \label{fig:muon}
\end{center}
\end{figure}

A High Level Trigger (HLT) is used to select events with at least one muon for both W and Z.  
The offline selection of ${\gamma^*/ \rm Z}\rightarrow \mu\mu$ candidates consists of the requirements of; two 
isolated\footnote{The isolation criteria requires
the $\pt$ sum of all tracks in a $\Delta R(=\sqrt{(\Delta \eta)^2 + (\Delta \phi)^2})$ cone of 0.3 around the muon
direction to be less than 3 GeV.} and opposite charged high $\pt$ ($>$20 GeV) muons measured in both the Tracker and
muon Chambers; invariant mass of the muons system, $M_{\mu \mu} > 40$ GeV. The resulting Z candidates for both signal
and backgrounds are shown in Fig.~\ref{fig:muon} as a function of $M_{\mu \mu}$. In order to reduce the 
QCD background significantly, the ${\rm W}\rightarrow \mu\nu$ events are selected with a higher $\pt$ threshold ($>$25 GeV)
and a tighter isolation criteria on muon candidates can be imposed. Events with two selected muon candidates are
rejected in order to reduce Z background to W. The Fig.~\ref{fig:muon} (left) shows that the QCD
background could be further reduced with a $M_{\rm T}>50~{\rm GeV}$ requirement. With these selections, 
~5500 ${\gamma^*/ \rm Z}\rightarrow \mu\mu$ candidates within 
$70<M_{\mu\mu}<140~{\rm GeV}$, and 64000 ${\rm W}\rightarrow \mu\nu$ selected 
events are expected in a data sample of $10~{\rm pb}^{-1}$ at $\sqrt{s}=14$ TeV~\cite{wz-mu}. 
CMS studies also the measurements of ${\rm Z}\rightarrow ee$ and ${\rm W}\rightarrow e\nu$ decays~\cite{wz-el}.

The efficiencies of reconstruction, isolation and trigger are measured in data using {\it tag-and-probe} methods. A sample
of ${\gamma^*/\rm Z}\rightarrow \mu\mu$ events with high purity is selected, where a higher $\pt$ and tighter isolation
requirements are imposed on one of the muons which acts as the event "tag". The efficiencies are measured with the
other muon which acts as the "probe". Furthermore, different methods to determine the QCD background  
in ${\rm W}\rightarrow \mu\nu$ using data are investigated. In the {\it matrix method} the background is estimated by
analyzing simultaneously two largely uncorrelated background-discriminating variables, which are in this case the
transverse mass of the W system and the isolation variable for muon. The systematic uncertainty due to 
the QCD background estimation with this method, is $0.4\%$. Other methods, e.g determining the $M_{\rm T}$ 
shape of W from Z events, transverse mass shape of the QCD background (to W) determination by reversing the isolation
criteria, are also investigated, and the results found to be consistent with the direct MC shapes at 
$<1\%$ level.

%\begin{figure}
%\begin{center}
%  \includegraphics[scale=0.30]{MET.eps}
%  \includegraphics[scale=0.30]{UpdateNew-ZeeNLO.eps} 
%  \caption{$\pt$ distribution of the Z boson in selected Z+$\ge 1$jet (left) and Z+$\ge 4$jet 
%  (right) for signal and background for an integrated luminosity of $1~fb^{-1}$.}
%  \label{fig:electron}
%\end{center}
%\end{figure}
%

\section{W(Z)+jets production in CMS}
The production of W(Z) bosons associated with jets at LHC has a wide range of physics potential, which varies 
from Standard Model (SM) measurements to Supersymmetry (SUSY) searches. 
These processes can be used for tests of perturbative quantum chromodynamic (QCD)~\cite{QCD}. 
The predictions for W(Z)$+N$Jets, where $N>2$, are accessible only 
through matrix element (ME) plus parton shower (PS) computations and in fact, 
can be considered as a prime testing ground for the accuracy of such predictions.
Z+jet events can be also exploited to calibrate jets, measured in the Calorimeter 
%, where Z boson momentum is balanced by one jet of hard subprocess 
(See~\cite{jetmet} for details). 
Furthermore, W(Z)+jets form a relevant background to many interesting phenomena, including new physics. Therefore, 
these processes must be measured with great accuracy to allow precision measurements and 
increase the sensitivity of the searches beyond SM.
%The cross section of this process is smaller than other calibration samples 
%(i.e. $\gamma$+jet), it rather provides a background free channel and the kinematics of 
%the Z boson can be reconstructed very precisely.

However, the individual cross section measurements
of W+Njets and Z+Njets will be affected by large systematic 
uncertainties associated mostly with the definition and measurement of jets. 
One of the measurements that CMS plans to perform is the ratio of the cross sections of W+jets to Z+jets 
as functions of jet multiplicity and boson $\pt$. 
%The measurement of the ratio, W$+N$jets/Z$+N$jets, 
Such a measurement allows partial cancellation 
of the most relevant experimental systematic uncertainties as well as the theoretical uncertainties due to
the choice of renormalization scale, the parton distribution functions, etc.~\cite{wzjets}. 
The jet energy scale forms the largest 
experimental uncertainty as it increases rapidly with jet multiplicity. This uncertainty 
cancels in the ratio as long as 
the $\pt$ spectra, the rapidity distribution and the 
composition of the jets in both processes are the same at the level of experimental sensitivity. 
Other uncertainties, i.e. Underlying Event (UE), Multiple interactions, 
luminosity and detector acceptances, will also cancel to a large
extent in the ratio.  
%The second largest uncertainty is expected to come from the Underlying Event (UE). 
%This contribution can promote a low energy jet above the 
%jet-counting threshold in $\pt$ and thus change the jet multiplicity of the event. 
%Studies show that the differences in W and Z production 
%are a small portion of the total and the uncertainties due to UE are negligible in the ratio~\cite{UE}. 
%
%Ratios can be used also for direct searches for new physics, such as large deviations from 1 in the 

%A number of physics generators are available to simulate major kinematic 
%properties of W(Z)+jets. 
The measurements of W(Z)+jets at the Tevatron collider 
indicate a general agreement between the theoretical predictions based on 
LO ME plus PS and data~\cite{CDF}.
In the studies presented here the ME event generator ALPGEN 
is used to generate exclusive parton level W(Z)+Njets (N=0,1,2,3,4,5) events. 
PYTHIA is used for PS and hadronization.   
The MLM recipe is used in order to avoid double counting of processes from ME and PS.  
The SM processes $t\bar{t}$+jets, WW+jets, WZ+jets, ZZ+jets and QCD
multi-jet are considered as backgrounds and generated with PYTHIA 
in fully inclusive decay modes for W and Z bosons. 

\begin{figure}
\begin{center}
  \includegraphics[scale=0.60]{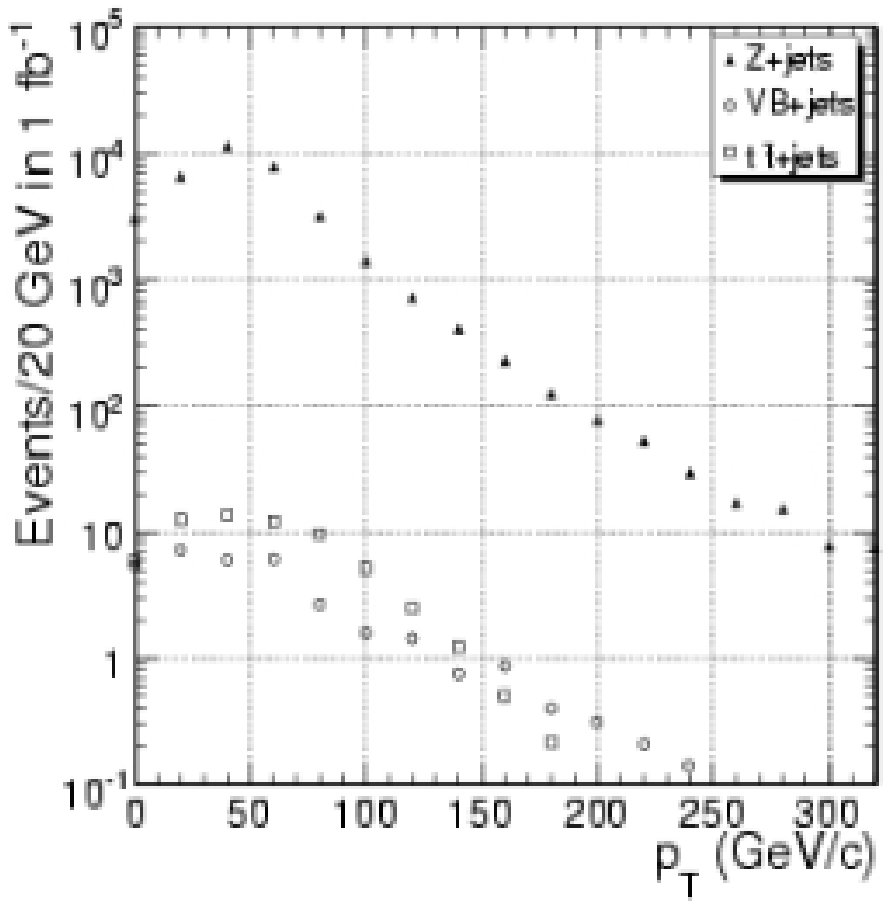}
  \includegraphics[scale=0.60]{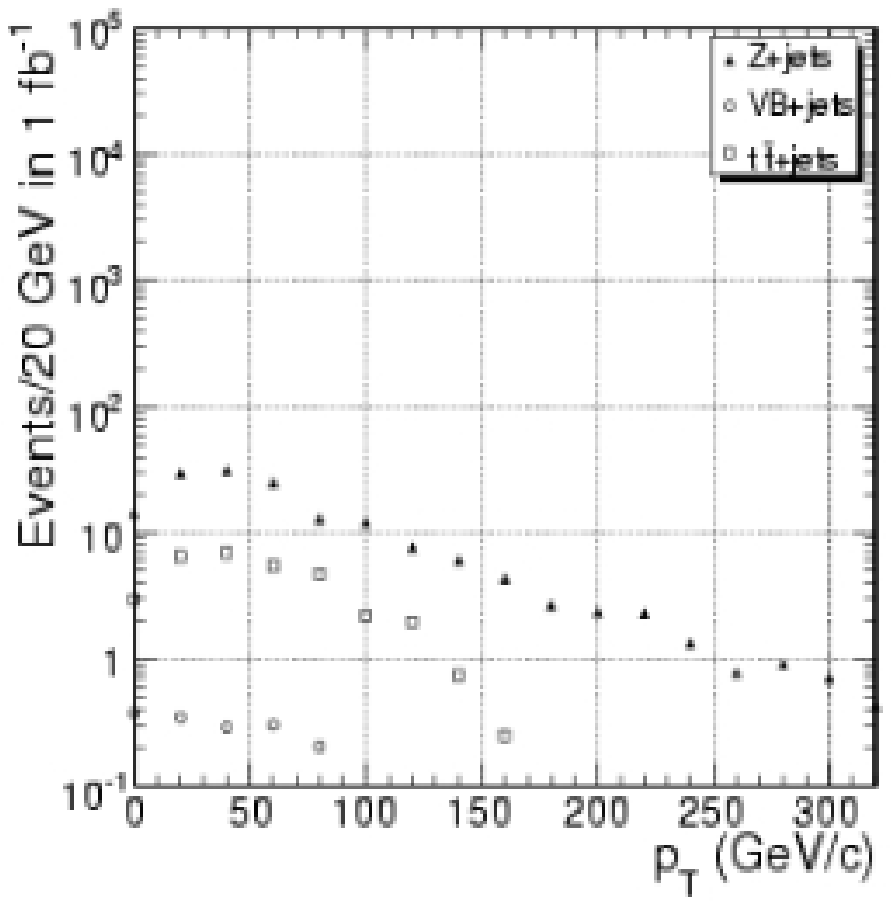} 
  \caption{$\pt$ distribution of the Z boson in selected Z+$\ge 1$jet (left) and Z+$\ge 4$jet 
  (right) for signal and background for an integrated luminosity of $1~fb^{-1}$.}
  \label{fig:zjets}
\end{center}
\end{figure}

\begin{figure}
\begin{center}
  \includegraphics[scale=0.60]{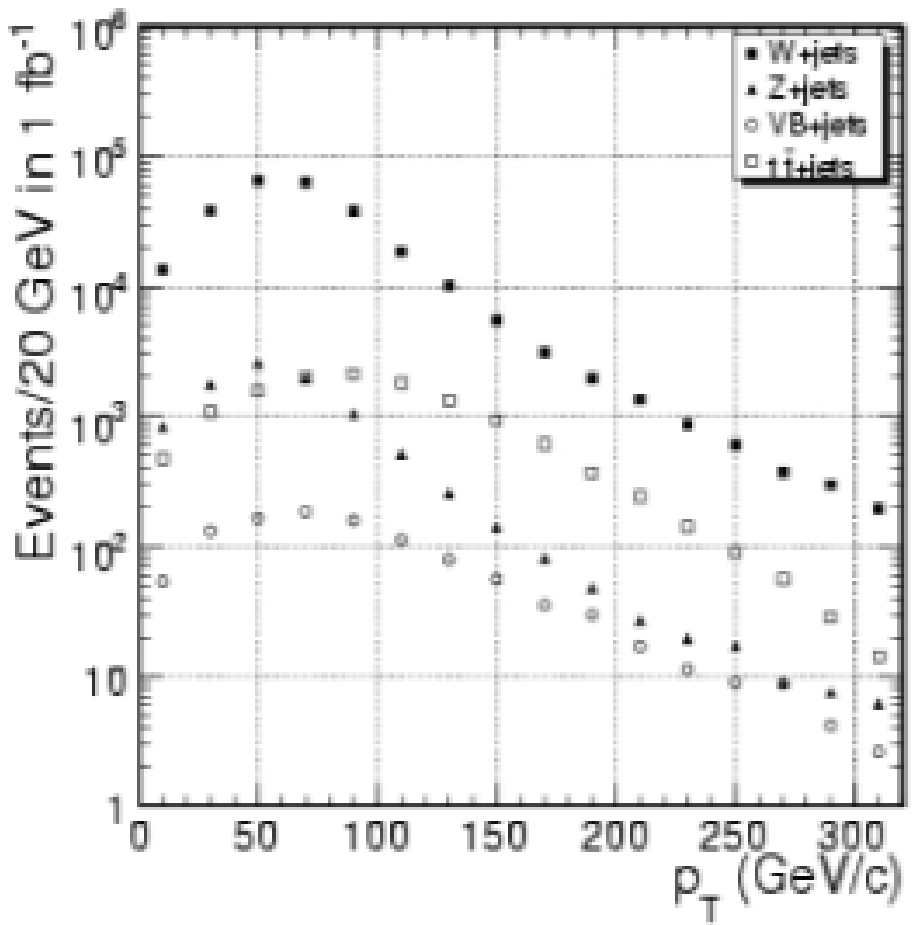}
  \includegraphics[scale=0.60]{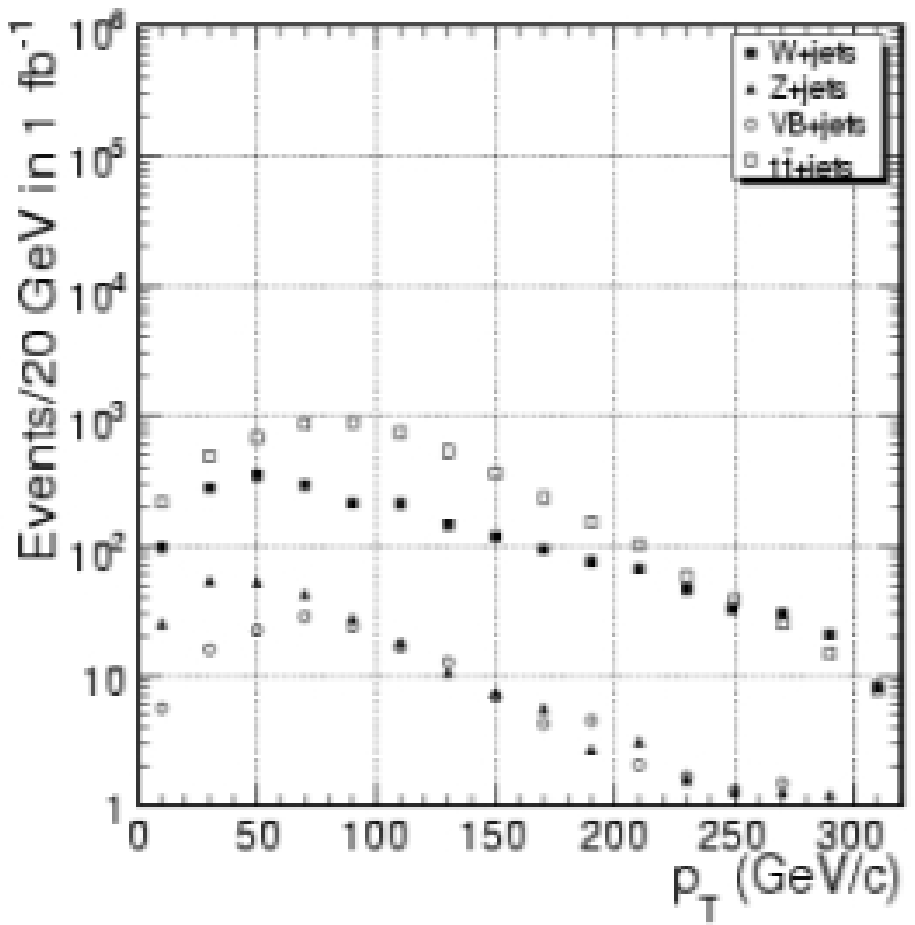} 
  \caption{$\pt$ distribution of the W boson in selected W+$\ge 1$jet (left) and W+$\ge 4$jet 
  (right) for signal and background for an integrated luminosity of $1~fb^{-1}$.}
  \label{fig:wjets}
\end{center}
\end{figure}

Figure~\ref{fig:zjets}(\ref{fig:wjets}) shows the $\pt$ distribution of the Z(W) boson in selected 
Z(W)+$\ge 1$jet (left) and Z(W)+$\ge 4$jet (right) events for the signal and backgrounds. 
In both W and Z boson cases 
the events are selected in  the electron and muon channels. The high $\pt$ isolated leptons are selected in 
order to reduce
contamination from QCD events. Furthermore, Z+jets events are selected by a tight di-lepton 
invariant mass around the
Z boson mass and the missing transverse energy is restricted to be small, whereas for 
W+jets a large missing transverse
energy is required. Jet reconstruction is performed using the Iterative Cone algorithm
using the energy deposited in the Calorimeter. Jets are calibrated using $\gamma$+jet events and the jets with 
$\pt > 50~ {\rm GeV}$ are counted. The current Z+jets selection provides a rather "clean" sample, and with 
$1~fb^{-1}$ of data, up to fourth jet multiplicity can be measured.   
One crucial point will be the reduction of the background to W$+>$2jets from $t\bar{t}$ events 
(See Fig.~\ref{fig:wjets} right), since the 
$t\bar{t}$ production rate increases by about a factor of 100 from the Tevatron to the LHC, while 
W production increases by just a factor of 5.  In the studies presented, the QCD contribution as 
background to W(Z)+jets is found to be negligible
and not shown in the figures. However, we should note that the background processes are 
simulated using the PYTHIA program,
which is known not to produce high jet multiplicities correctly.
%The data driven methods for background estimation are therefore
%extremely important.
%
\section{Conclusions}
CMS has established a strategy for early measurements of the inclusive W and Z production cross section
with a data of $10~pb^{-1}$. The data-driven methods for background estimation, measuring the
efficiencies are studied.  
The measurements of W(Z)+jets versus 
the jet multiplicity will be one of the early measurements carried out with the CMS detector and the analysis
strategy will be adapted (to large extent) from the inclusive W(Z) measurement. 
The ratio measurement of W+Njets to Z+Njets will allow partial cancellation of the most relevant 
systematic uncertainties. This is an extremely important advantage at the startup, where it will be 
relatively difficult to control the detector related 
systematic uncertainties. 

%The ratio measurement can benefit from other types of jets 
%(e.g. reconstructed with Tracker Tracks only) than the standard Calorimeter
%based jets. Unlike Calorimeter jets Track based jets preserve the vertex information 
%which helps to count jets originating only 
%from the signal vertex, eliminating Pile Up contamination to the jets.  
%


\begin{thebibliography}{999}
\bibitem{wz-mu}
CMS Collaboration, %{\it Towards a Measurement of the Inclusive ${\rm W}\rightarrow \mu\nu$ and 
%${\rm Z}\rightarrow \mu\mu$ cross sections in pp collisions at $\sqrt{s}=14~{\rm TeV}$}, 
PAS 2007/002 (2008).\\[-7mm]
\bibitem{wz-el}
CMS Collaboration, 
%{\it Towards a Measurement of the Inclusive ${\rm W}\rightarrow e\nu$ and 
%$\gamma^*/ {\rm Z}\rightarrow {\rm e^{+}e^{-}}$ cross sections in pp collisions at $\sqrt{s}=14~{\rm TeV}$}, 
PAS EWK-08-005 (2008).\\[-7mm]
%
\bibitem{QCD} J. R. Andersen {\it et al.}, 
%{\it W Boson Production with Associated Jets at Large Rapidity}, 
JHEP 05 (2001) 048, arXiv:hep-ph/0105146v115.\\[-7mm]
%
%\bibitem{iki}J. M. Campbell et al, {\it Hard interactions of quarks and gluons: a primer for LHC physics}, 
%Rep. Prog. Phys. 70 89-193, (2007).
%
\bibitem{jetmet}
D. Dobur [On behalf of the CMS Collaboration], {\it Jets and Missing Transverse Energy Reconstruction in CMS}, 
to be published in the proceedings of the {\it Physics at LHC, SPLIT08} conference 
(September 29 - 4 October, 2008).\\[-7mm]
%
\bibitem{wzjets} E. Abouzaid and H. Frisch, 
%{\it The Ratio of W+Njets To Z/$\gamma $+Njets As a Precision Test of the Standard Model},
arXiv:hep-ph/0303088v1, (2003).\\[-7mm]
%
%\bibitem{UE} T. Affolder {\it et al.} (CDF Collaboration), 
%{\it Charged jet evolution and the underlying event in proton-antiproton collisions at 
%1.8 TeV}, Phys.Rev.D 65, 092002 (2002). 
%%
\bibitem{CDF} T. Aaltonen {\it et al.} [CDF Collaboration], 
%{\it Measurement of the cross section for W-boson production in association with jets in 
%ppbar collisions at sqrt(s) = 1.96 TeV},
 Phys. Rev. D 77, 011108(R)(2008).\\[-7mm]
%

\end{thebibliography}
\end{document}